\documentclass[reprint,amsmath, amssymb, aps, floatfix,
]{revtex4-2}
\usepackage{graphicx}
\usepackage{dcolumn}
\usepackage{bm}
\usepackage[usenames,dvipsnames]{color}
\usepackage[svgnames]{xcolor}
\usepackage{float}
\usepackage{amsmath}

\begin{document}

\preprint{APS/123-QED}

\title{Survival of black holes through a cosmological bounce }

\author{Daniela P\'erez}
\email{danielaperez@iar.unlp.edu.ar}

 \affiliation{%
 Instituto Argentino de Radioastronom\'ia (IAR, CONICET/CIC/UNLP), C.C.5, (1894) Villa Elisa, Buenos Aires, Argentina}%



\author{Gustavo E. Romero}
\altaffiliation[Also at ]{Facultad de Ciencias Astron\'omicas y Geof\'isicas, Universidad Nacional de La Plata, Paseo del Bosque s/n, 1900 La Plata, Buenos Aires, Argentina}
\affiliation{%
Instituto Argentino de Radioastronom\'ia (IAR, CONICET/CIC/UNLP), C.C.5, (1894) Villa Elisa, Buenos Aires, Argentina }%


\date{\today}

\begin{abstract}
We analyze whether a black hole can exist and survive in a universe that goes through a cosmological bounce. To this end, we investigate a central inhomogeneity embedded in a bouncing cosmological background modeled by the comoving generalized McVittie metric. Contrary to other dynamical metrics available in the literature, this solution allows for the interaction of the central object with the cosmological fluid. We show that the horizons associated with this metric change with cosmic time because they are coupled to the cosmic evolution as the mass of the central object is always proportional to the scale factor: it decreases during contraction and increases during expansion phases. After a full analysis of the causal structure of this spacetime, we determine that a dynamical black hole persists during the contraction, bounce, and expansion of the universe. This result implies that there is a class of bouncing models that admits black holes at all cosmological epochs. If these models are correct approximation to the real universe, then black holes surviving a cosmic collapse could play some role in the subsequent expanding phase.

\end{abstract}

\maketitle


\section{Introduction}\label{sec:1}

The standard model of cosmology is based on Einstein field equations with a cosmological constant term, the validity of the standard model of quantum field theory, a hot and dense phase at the beginning of the cosmic expansion, and the hypothesis that the content of the universe is a perfect fluid plus a dominant cold dark matter (CDM) component that only manifests itself gravitationally. An initial exponentially expanding phase is usually added to explain the existence of primordial inhomogeneities and other observational features. This model, called $\Lambda$CDM, works very well to explain the available data \cite{pla+20,Ade:2015xua}.  It is, nonetheless, not free of problems. 

One of the important tensions of the $\Lambda$CDM model is the incompatibility of the current model of particle physics with the existence of dark matter. Extensions of the model must be implemented to predict candidates for dark matter, and such extensions are not unique. Moreover, all experiments attempting at detecting dark matter particles have failed so far.  Another problem of the standard cosmology is the lack of understanding of the inflaton, the presumed scalar field responsible for the initial cosmic inflation. 

Additionally, there is a $4\sigma$ to $6\sigma$ disagreement between the locally measured expansion rate of the universe, quantified by the Hubble constant, and the value inferred from the cosmic microwave background measurements by Planck satellite in the context of the $\Lambda$CDM model. This problem is known in cosmology as ``the Hubble tension'', and might be a strong indication of the inadequacy of the $\Lambda$CDM model (for a recent and comprehensive review on this problem see \cite{div+21}).

A major problem is that the $\Lambda$CDM model is singular. A singularity is an undesirable feature of any physical description of nature because it implies the incompleteness of the background theory \cite{rom13}, in this case general relativity. One approach frequently adopted to avoid the singularity problem is to invoke a quantum theory of gravity that presumably would lack the undesirable features and would rule at the Planck scale. But there is no clear picture of quantum gravity yet. Much less a consistent theory of it. A different approach is to avoid the singularity through a classical cosmological bounce from a previous contracting phase. During the contraction, the density and temperature of the cosmic fluid increases in such a way that all structure is expected to be erased. It is not clear, however, whether black holes formed before the bounce would survive to it. 

If black holes get through the bounce, they can produce perturbations that would give rise to structure and early galaxy formation in the expanding phase \cite{car+18}. Persistent black holes might also form the dark matter without need of invoking new particle physics \cite{car+20,gre+21}. Black holes from the bounce can even help to explain recently observed LIGO/VIRGO events with inferred black hole masses well inside the pair-instability gap \cite{jed20}. 

The survival of black holes to a cosmological bounce, however, is unclear. Carr and Cooley \cite{car+11} presented a first semi-qualitative discussion of the problem.  They noticed that accretion might play an important role for black holes getting through the bounce. Clifton et al. \cite{Clif+2017} investigated the behavior of a lattice of black holes in a universe whose energy density
is dominated by a scalar field and goes from a Big Crunch to a Big Bang. They obtained some exact solutions for time-evolving models in which multiple distinct black holes persist through the bounce. More recently, Gorkavyi and Tyul'bashev \cite{gor+21} investigated the effects of multiple bounces on a population of non-dynamical black holes. 

Since black holes are essentially regions of spacetime, the global evolution of the universe should affect their horizons, especially close to a bounce. In a previous paper we considered the evolution of the McVittie metric before, during, and after a cosmic bounce \cite{per+21a}. We showed that although the metric describes a black hole in the past of the bounce, the trapping horizon disappears close to it, when it merges with the cosmic horizon. This result remains valid even when the bounce is not symmetric \cite{per+21b}.  If the black hole interacts with the cosmic fluid, however, the McVittie metric is not an adequate description of the situation because in such metric the central mass remains constant. In the present paper we deal with this problem and we investigate a black hole described by a generalized McVittie metric. These black holes interact with the cosmic fluid and and evolve with the universe.

\section{Scale factor bouncing cosmological model}

In a bouncing cosmology the universe contracts from a very diluted phase. The contraction then smoothly evolves into a bounce that leads to the current phase of expansion as described by the $\Lambda$CDM model. By construction, the cosmological singularity is absent in such bouncing models. 

There are many mechanisms that could generate a cosmological bounce, either by classical \cite{Wands:2008tv, Ijjas:2016tpn, Galkina:2019pir} or quantum \cite{Peter:2008qz, Almeida:2018xvj, Bacalhau:2017hja, Frion:2018oij}  effects. The reader is referred to the review by Novello and Bergliaffa \cite{Novello2008} for further details about these mechanisms.

We choose a scale factor that provides a bounce, has a simple analytical form, and describes a realistic cosmological model \cite{Frion2020,pet+07}:
 \begin{equation}\label{scale-factor}
 a(T) = a_b \left[1 + \left(\frac{T}{T_b}\right)^2\right]^{1/3}.
  \end{equation}
This scale factor was obtained by Peter and collaborators \cite{pet+07} considering quantum corrections to the classical evolution of the scale factor. The corrections were obtained by solving the Wheeler-de Witt equation in the presence of a single perfect fluid, within the framework of the de Broglie-Bohm quantum theory \cite{Pinto-Neto:2013toa}. 


In this model, the bounce occurs due to quantum cosmological effects when the curvature of spacetime becomes very large. There is no phantom field that causes the bounce.
  

The lower limit on $T_b$, given by $T_b > 10^{3} t_{\mathrm{Planck}}$, is set to ensure the validity of the Wheeler-de Witt equation, which was employed to derive the scale factor \eqref{scale-factor} \cite{Frion2020}.


The upper bound is set in order for the bounce to occur before the Big Bang Nucleosynthesis, as explained in \cite{Frion2020}. Thus, $10^{3} \; t_{\mathrm{Planck}} < T_{b} < 10^{40} \; t_{\mathrm{Planck}}$, i.e. $10^{-41} \; \mathrm{s} < T_{b} < 10^{-4} \; \mathrm{s}$. We adopt a value close to the upper limit ($T_{b} =10^{-4} \; \mathrm{s}$).
  
We see from Eq. \eqref{scale-factor} that for $T >>T_b$ the scale factor reduces to that of dust. Near the bounce the evolution is driven by an effective fluid with negative energy density that scales as $a^{-6}$, as can be seen from Friedman's equation\footnote{The quantum effect which causes the bounce could be equivalently obtained from an effective Friedmann equation (adding to the Friedmann equation a term corresponding to a stiff matter fluid with negative energy density):
\begin{equation}
\left(\frac{\dot{a}}{a}\right) = \frac{8 \pi G}{3 c^{2}}\rho \left(1 - \frac{\rho}{\rho_c}\right),
\end{equation}
where $\rho_c$ is the critical density at $T_b$ and $\rho_c \approx c^{2}/(24 G T^{2}_b$.}.
The violation of the energy conditions around $T_{b}$ assures the non-applicability of the singularity theorems \cite{Hawking-Ellis-book}.


\section{Generalized McVittie spacetime: comoving case}

The generalized McVittie spacetime, originally proposed by Faraoni and Jacques \cite{far+07}, has the following line element written in isotropic coordinates $(T, r, \theta, \phi)$ \cite{far+07, far15} 
\begin{eqnarray}\label{GMcV}
 ds^{2} & = & - \frac{\left[1 - \frac{G m(T)}{2 c^2 r}\right]^{2}}{\left[1 + \frac{G m(T)}{2 c^2 r}\right]^{2}} c^2 dT^{2}\\
 & + & a(T)^{2} \left[1 + \frac{G m(T)}{2 c^2 r}\right]^{4} \left[dr^{2} + r^{2}\left(d\theta^{2} + \sin^{2}{\theta} d\phi^{2}\right)\right].\nonumber
\end{eqnarray}
In this expression, $a(T)$ is the scale factor of the cosmological background model and $m(T)$ is a function that depends on the cosmic time $T$. The only non-vanishing mixed components of the Einstein tensor are \cite{far+07,far15}
\begin{align}
{G^{T}}_{T} & =  - 3 \frac{A(T,r)^2}{B(T,r)^2} C(T,r)^{2},\\
{G^{r}}_{T} & =  \frac{2G \, m(T)}{c^{2} r^{2} a(T)^2 B(T,r) A(T,r)^5} \left(\frac{\dot{a(T)}}{a(T)} +\frac{\dot{m(T)}}{m(T)} \right),\\
{G^{r}}_{r} & =  {G^{\theta}}_{\theta}  =  {G^{\phi}}_{\phi} = - \frac{A(T,r)^2}{B(T,r)^2}  D(T,r).
\end{align}
where
\begin{align}
A(T,r) & =    \left[1 + \frac{G\,  m(T)}{2 c^2 r}\right],\\
B(T,r) & =  \left[1 - \frac{G \, m(T)}{2 c^2 r}\right],\\
C(T,r) & =  \left[\frac{\dot{a(T)}}{a(T)}+ \frac{G}{c^2 r}\frac{\dot{m(T)}}{A(T,r)}\right],\\
D(T,r) & =  2 \dot{C}(T,r) \nonumber \\
& +  C(T,r) \left(3 C(T,r) + \frac{2 G \dot{m}(T)}{c^{2} r A(T,r) B(T,r)}\right).
\end{align}

When solving the Einstein field equations with an energy-momentum tensor of a perfect fluid, the only possible solutions are the Schwarzschild-de Sitter metric or the McVittie metric (see, for instance, \cite{car+09}). Thus, we consider an imperfect fluid with energy-momentum tensor given by
\begin{equation}\label{energy-momentum}
 T_{ab} = \left(\frac{p}{c^{2}}+\rho\right) u_{a} u_{b} + p g_{ab} + q_{a} u_{b} + q_{b} u_{a}. 
\end{equation}
Here, $\rho$ is the density, $p$ is the pressure, $u^{a}$ is the four-velocity of the fluid, and $q^{a}$ is a spatial vector field that represents the current density of heat. If we assume that
\begin{equation*}
u^{\mu} = \left(\frac{A(T,r)}{B(T,r)},0,0,0\right), \;\;\; q^{\alpha} = \left(0,q,0,0\right),\;\;\;u^{b} q_{b} =0,  
\end{equation*}
then Einstein field equations reduce to 
\begin{eqnarray}
\frac{\dot{a(T)}}{a(T)} +\frac{\dot{m(T)}}{m(T)} & = & - \frac{4 \pi r^2 a^2(T)}{m(T)} B^{2}(T,r) A^4(T,r) q,\label{G10}\\
\rho(T,r) & = & \frac{3 c^{2}}{8 \pi G} \frac{A(T,r)^2}{B(T,r)^2} C(T,r)^{2},\label{density}\\
p(T,r) & = & - \frac{c^{4}}{8 \pi G}  \frac{A(T,r)^2}{B(T,r)^2}  D(T,r). \label{pressure}
\end{eqnarray}
This is an underdetermined system with 3 equations for 5 unknown functions: $a(T)$, $m(T)$, $\rho(T,r)$, $p(T,r)$ and $q$. A possible way of arriving at a particular solution is by specifying the functions $a(t)$ and $m(T)$, and then solving the equations for $\rho(T,r)$, $p(T,r)$ and $q$ \cite{car+09}. We shall follow this path.

Notice that if $q = 0$, the energy-momentum tensor given by \eqref{energy-momentum} corresponds to a perfect fluid and Eq. \eqref{G10} becomes
\begin{equation}
 \frac{\dot{a(T)}}{a(T)} +\frac{\dot{m(T)}}{m(T)} = 0.
\end{equation}
The solution is
\begin{equation}\label{condMcV}
m(T) = \frac{m_0}{a(T)},  
\end{equation}
where $m_0$ is a non-negative constant. If we substitute expression \eqref{condMcV} into the line element \eqref{GMcV} we arrive at the McVittie metric. The constant $m_0$ can be interpreted, in the Newtonian limit, as the mass of the central object \footnote{The reader is referred to Carrera and Giulini (2010) \cite{car+10}, Appendix 3 for further details.}.

In spherically symmetric spacetimes, the Misner-Sharp-Hernandez (MSH) energy \cite{mis+64} provides a good measure of the local mass-energy of the system. It is defined in terms of the Riemann curvature which allows a decomposition into a sum of two terms corresponding to the Ricci and Weyl curvature, respectively. The Ricci part locally relates to the energy-momentum tensor while the Weyl part identifies the gravitational energy of the central object.

Carrera and Giulini \cite{car+09} computed the Weyl part of the MSH energy for the generalized McVittie metric
\begin{equation}\label{weyl-part}
E_{\mathrm{W}} = a(T) m(T).  
\end{equation}
McVittie spacetime ($m(t) = m_0/a(T)$) leads to $E_{\mathrm{W}} = m_0$, that is, the gravitational energy of the central object remains constant. On the contrary, in the generalized case, the 
Weyl part becomes dependent on time suggesting a connection with cosmic dynamics.


In what follows, we focus on a particular class of generalized McVittie solutions that corresponds to the choice 
\begin{equation} \label{mt}
m(T) = m_{0},
\end{equation}
where $m_0$ is a constant, and the corresponding Weyl part of the MSH mass is (see \eqref{weyl-part})
\begin{equation}
E_{\mathrm{W}} = m_0 a(t).  
\end{equation}
Thus, we see that the central object is coupled with the cosmic evolution.

If we substitute Eq. \eqref{mt} into the line element \eqref{GMcV}, we obtain
\begin{eqnarray}\label{GMcVS}
ds^{2} & = & - \frac{\left[1 - \frac{G m_0}{2 c^2 r}\right]^{2}}{\left[1 + \frac{G m_0}{2 c^2 r}\right]^{2}} c^2 dT^{2}\\& + & a(T)^{2} \left[1 + \frac{G m_0}{2 c^2 r}\right]^{4} \left[dr^{2} + r^{2}\left(d\theta^{2} + \sin^{2}{\theta} d\phi^{2}\right)\right].\nonumber
 \end{eqnarray}
In the limit $a(T) \rightarrow 1$, the Schwarzschild metric in isotropic coordinates is recovered, and if $m_{0} \rightarrow 0$, we obtain the Friedmann-Lema\^{i}tre-Robertson–Walker (FLRW) cosmological spacetime.

The expressions of the Kretschmann ($K$), Weyl ($W$), and Ricci ($R$) scalars of the comoving generalized McVittie (CGMcV) metric are
\begin{equation}
K = 4 \frac{K_1+K_2}{K_3},  
\end{equation}
where
\begin{eqnarray}
 K_1 & = &  -1024 c^{14} G^2 {m_0}^2 r^4 a'(T)^2 \left(2 c^2 r+G m_0\right)^8 \nonumber \\
 & + & 3 a'(T)^4 \left(2 c^2 r+G m_0\right)^{16},\\
 K_2 & = & 3 a(T)^2 a''(T)^2 \left(2 c^2 r+G m_0\right)^{16} \nonumber \\
 & + & 49152 c^{24} G^2 {m_0}^2 r^6 \left(G m_0-2 c^2 r\right)^4,\\
 K_3 & = & c^4 a(T)^4 \left(G m_0-2 c^2 r\right)^4 \left(2 c^2 r+G m_0\right)^{12},
\end{eqnarray} 
\begin{eqnarray}
W & = & \frac{196608 \; c^{20} G^2 {m_0}^2 r^6}{a(T)^4 \left(2 c^2 r+G m_0\right)^{12}},
\end{eqnarray}
and
\begin{eqnarray}
R & = & \frac{6 \left(a'(T)^2+a(T) a''(T)\right) \left(2 c^2 r+G m_0\right)^2}{c^2 a(T)^2 \left(2 c^2 r-G m_0\right)^2}.
\end{eqnarray}

Both the Kretschmann and Ricci scalar are singular at $r = G m_0/2 c^{2}$ while the Weyl scalar is perfectly defined everywhere. The Ricci scalar is associated with the behaviour of the matter since $R = - 8 \pi G / c^{4} \; T$, where $T \equiv {T^{\mu}}_{\mu}$ is the trace of the energy-momentum tensor\footnote{The expression $R = - 8 \pi G / c^{4} \; T$ can be derived from Einstein field equations written in terms of the mixed components
\begin{equation}
{R^{\mu}}_{\nu} - \frac{1}{2} {\delta^{\mu}}_{\nu}  R = \frac{8 \pi G}{c^{4}} {T^{\mu}}_{\nu}.
\end{equation}
If we contract and set $\mu = \nu$, we immediately obtain $R = - 8 \pi G / c^{4} \; T$. }. Because the Weyl scalar is smooth and continuous through $r = G m_0/2 c^{2}$, this surface does not constitute an essential singularity of the spacetime. The divergence in the Ricci scalar is thus pointing out some anomaly in the matter properties in that particular region as we show below.

Under the choice $m(T) = m_0$, the energy density and pressure (see Eqs. \eqref{density} and Eqs. \eqref{pressure}) are
\begin{eqnarray}
\rho(T,r) & = & \frac{3 c^{2}}{8 \pi G} \frac{\left[1 + \frac{G\,  m_0}{2 c^2 r}\right]^{2}}{ \left[1 - \frac{G \, m_0}{2 c^2 r}\right]^{2}} H^{2}(T),\label{density-1}\\
p(T,r) & = & - \frac{c^{4}}{8 \pi G} \frac{\left[1 + \frac{G\,  m_0}{2 c^2 r}\right]^{2}}{ \left[1 - \frac{G \, m_0}{2 c^2 r}\right]^{2}} \left[2 \dot{H}(T) + 3 H^{2}(T)\right].\label{pressure-1}
\end{eqnarray}
In the limit $m_0 \rightarrow 0$, we recover the expressions for the energy density and pressure in FLRW spacetime. In terms of the scale factor \eqref{scale-factor}, expressions \eqref{density-1} and \eqref{pressure-1} take the form
\begin{align}
\rho(T,r) & =  \frac{ c^{2}}{6 \pi G} \frac{\left[1 + \frac{G\,  m_0}{2 c^2 r}\right]^{2}}{ \left[1 - \frac{G \, m_0}{2 c^2 r}\right]^{2}} \frac{1}{T^{2}_{b}}\left(\frac{T}{T_b}\right)^{2} \frac{1}{\left[1 + \left(\frac{T}{T_b}\right)^{2}\right]^{2}},\label{density-2}\\
p(T,r) & =  - \frac{c^{4}}{6 \pi G} \frac{\left[1 + \frac{G\,  m_0}{2 c^2 r}\right]^{2}}{ \left[1 - \frac{G \, m_0}{2 c^2 r}\right]^{2}} \frac{1}{T^{2}_{b}} \frac{1}{\left[1 + \left(\frac{T}{T_b}\right)^{2}\right]^{2}}.\label{pressure-2}
\end{align}

The energy density is always positive, except at $r = G m_0/2 c^{2}$ where it is not defined \footnote{The energy density evaluated at $T = T_{b}$, and ignoring the spatial part, yields $\rho(T_b) \approx c^{2}/(24 \pi G T^{2}_{b})$ that for $T_{b} = 10^{-4}$ s is of the order of $ \approx 10^{33}$ erg, or $ \approx 10^{13}$ g/$\mathrm{cm}^{3}$. The mechanism that produces the bounce at such nuclear densities was fully developed in Peter et al. \cite{pet+07}}. The pressure is also divergent on this surface. Since the Weyl scalar is perfectly defined on the surface, the presence of pathologies in the Kretschmann and Ricci scalars might not necessarily imply geodesic incompleteness but the inadequacy of the matter model adopted.

The pressure of the fluid is always negative, with a minimum value at the bounce. For large positive and negative values of the cosmic time, it goes to zero, that is, the equation of state is that of dust-like matter, which is in accordance with the scale factor \eqref{scale-factor} for $T >>T_b$.

The violation of the strong energy condition in the context of General Relativity ensures the existence of a cosmological bounce, as discussed in \cite{Novello2008}. In the present model, there is a time interval for which $\rho c^{2}+ p \le 0$:
\begin{equation}
  \rho c^2 + p  \le 0   \Rightarrow  \left(\frac{T}{T_b}\right)^{2} -1  < 0, 
   \forall T, T \in \left[- T_{b}, T_{b}\right].
  \end{equation}
Thus, the constant $T_{b}$ provides the timescale where the strong energy condition is violated.

In terms of the Weyl part of the MSH energy, Eq. \eqref{G10} yields
\begin{equation}
q = - \frac{\dot{E}_{W}}{4 \pi r^{2} a^{3}(T) B^{2}(T,r) A^{4}(T,r)}.  
\end{equation}
If $q >0$ ($q <0$), for heat flowing in an outward (inward) pointing radial direction, then the Weyl part decreases (increases). In the contracting phase previous to the bounce, the scale factor diminishes and $q >0$, while in the expanding phase $q < 0$ and the Weyl part grows along with the scale factor.

If the solution \eqref{GMcVS} describes a black hole, as we prove in Sec. \ref{CGMcVPG}, the coupling of the black hole with the cosmic dynamics implies that the black hole loses mass until the bounce. Afterwards, in the expanding phase, there is a heat flux toward the black hole and its mass increases. This behaviour is imposed by the coupling between the central object and the cosmological evolution in the present model, but it is does not take into account the accretion process on local scale, that can change the mass change rate.



 For the upcoming analysis, it is convenient to express \eqref{GMcVS} in terms of the radius coordinate $\tilde{r}$, that is related to the isotropic radius $r$ by 


\begin{equation}
 \tilde{r} \equiv r \left(1 + \frac{G m_{0}}{2 c^{2} r}\right)^{2},
\end{equation}
or equivalently
\begin{equation}
 r = \frac{\tilde{r}}{2}\left[1 - \frac{G m_0}{c^{2}\tilde{r}} \pm \sqrt{1 - \frac{2 G m_0}{c^{2}\tilde{r}}}\right]. 
\end{equation}
The latter relation makes apparent that $r = r(\tilde{r})$ is not a bijective function; the isotropic radius $r$ maps twice the spacetime region $\tilde{r} > 2 G m_0/c^{2}$, and does not cover the range $0 < \tilde{r} < 2 G m_0/c^{2}$.

By making the coordinate transformation $(T,r,\theta,\phi) \; \rightarrow \; (T,\tilde{r},\theta,\phi)$, the line element  \eqref{GMcVS} takes the form
\begin{eqnarray}\label{schwcosmo}
 ds^{2} & = & - c^{2} \left(1 -  \frac{2 G m_{0}}{ c^{2} \tilde{r}}\right) dT^{2}\\
 & + & a^{2}(T)\left(1 -  \frac{2 G m_{0}}{ c^{2} \tilde{r}}\right)^{-1} {d\tilde{r}}^{2} + a^{2} {\tilde{r}}^{2}\left(d\theta^{2} + \sin^{2}{\theta} d\phi^{2}\right).\nonumber
\end{eqnarray}

The CGMcV metric has been previously analyzed \cite{gao+08,far15} in terms of the areal radius coordinate $R = a(T) \tilde{r}$. Under the coordinate transformation $(T,\tilde{r},\theta,\phi) \;  \rightarrow \; (T,R,\theta,\phi)$ Eq. \eqref{schwcosmo} yields
\begin{eqnarray}\label{ds-areal-radius}
ds^{2} & = & -c^2 \left(f(T,R) - \frac{ H(T)^2 R^2}{c^2 f(T,R)}\right) dT^2 + \frac{dR^2}{f(T,R)}  \nonumber \\
& - &  2 \frac{H(T) R}{f(T,R)} dT dR +  R^2   \left(d\theta^{2} + \sin^{2}{\theta} d\phi^{2}\right),
\end{eqnarray}
where,
\begin{equation}
 f(T,R) = 1 - \frac{2 G m_0 a(T)}{c^2 R} ,
\end{equation}
and $H(T) \equiv \dot{a}/a$ is the Hubble factor. In what follows, we argue that these coordinates are not adequate to unveil certain important features of this metric; instead, we present an alternative set of coordinates that are more suitable to our purpose. 

We first compute the trapping horizons. Theses are defined as the surfaces where null geodesics change their focusing properties \cite{hay94}. Mathematically, these horizons are determined by the condition
\begin{equation}
 \theta_{\mathrm{in}} \theta_{\mathrm{out}} = 0, 
\end{equation}
where $\theta_{\mathrm{in}}$ stands for the expansion of ingoing radial null geodesics with tangent field $n^{a}$, whereas $\theta_{\mathrm{out}}$ denotes the expansion of outgoing radial null geodesics with tangent field $l^{a}$, respectively. Spacetime regions can be classified as:
\begin{itemize}
\item   \textit{Regular} if $\theta_{\mathrm{in}} \theta_{\mathrm{out}} < 0$.
\item  \textit{Anti-trapped} if  $\theta_{\mathrm{in}} \theta_{\mathrm{out}} > 0$, where $\theta_{\mathrm{in}} > 0$ and $\theta_{\mathrm{out}} >0$.
\item \textit{Trapped} if $\theta_{\mathrm{in}} \theta_{\mathrm{out}} > 0$,  where $\theta_{\mathrm{in}} < 0$ and $\theta_{\mathrm{out}} < 0$.
\end{itemize}
Trapped regions are a key feature that allow to identify the presence of a black hole \cite{hay94}: in the trapped region of a black hole ingoing and outgoing null rays are converging and remain confined and enclosed by a horizon.

The expansion of the null vector $l^{a}$ when the geodesic to which it is tangent is not necessarily affinelly-parametrized can be computed using the expression \cite{far15}
\begin{equation}\label{theta-out}
  \theta_{\mathrm{out}} = \left[g^{ab} + \frac{l^{a} n^{b} + n^{a} l^{b}}{\left(-n^{c} l^{d} g_{cd}\right)}\right] \nabla_{a}l_{b}.
\end{equation}
In the same way, the expansion of the null vector $n^{a}$ can be determined using the relation
\begin{equation}\label{theta-in}
  \theta_{\mathrm{in}} = \left[g^{ab} + \frac{l^{a} n^{b} + n^{a} l^{b}}{\left(-n^{c} l^{d} g_{cd}\right)}\right] \nabla_{a}n_{b}.
\end{equation}
Though both $\theta_{\mathrm{out}}$ and $\theta_{\mathrm{in}}$ are scalar quantities, they depend on the choice of the vectors $l^{a}$ and $n^{a}$. Each pair of these vectors defines a set of spacelike surfaces normal to them. In other words, trapping horizons depends on the election of the spacetime foliation, i.e. on the time coordinate \cite{nil+11}.

Because of the spherical symmetry, the equation for the ingoing and outgoing radial null geodesics can be derived by setting $d\theta = d\phi = 0$ in $ds^{2} = 0$, thus obtaining
\begin{equation}\label{out-in}
\left. \frac{dR}{dT} \right\vert_{\pm} =  H(T) R \pm  c f(T,R),
\end{equation} 
 where the ``$-$'' (``$+$'') corresponds to the ingoing (outgoing) case. The tangent vector fields $l^{a}$ and $n^{a}$ have the form
 \begin{eqnarray}
 n^{\mu} & = & \left(1, - c f(T,R) + H(T) R, 0, 0\right),\\
 l^{\mu} & = & \left(1, c f(T,R) + H(T) R, 0, 0\right).
 \end{eqnarray}

Finally, the expansions $\theta_{\mathrm{out}}$ and $ \theta_{\mathrm{in}}$ yield:
\begin{eqnarray}
 \theta_{\mathrm{out}} & = &  \frac{2 \ c}{R}\left[ 1 - \frac{2 G m_0 a(T)}{c^2 R} + \frac{R H(T)}{c}\right], \label{gout}\\
 \theta_{\mathrm{in}} & = & \frac{2 \ c}{R}\left[- 1 + \frac{2 G m_0 a(T)}{c^2 R} + \frac{R H(T)}{c}\right].\label{gin}
 \end{eqnarray}
 
\begin{figure}[t]
\includegraphics[width=8cm]{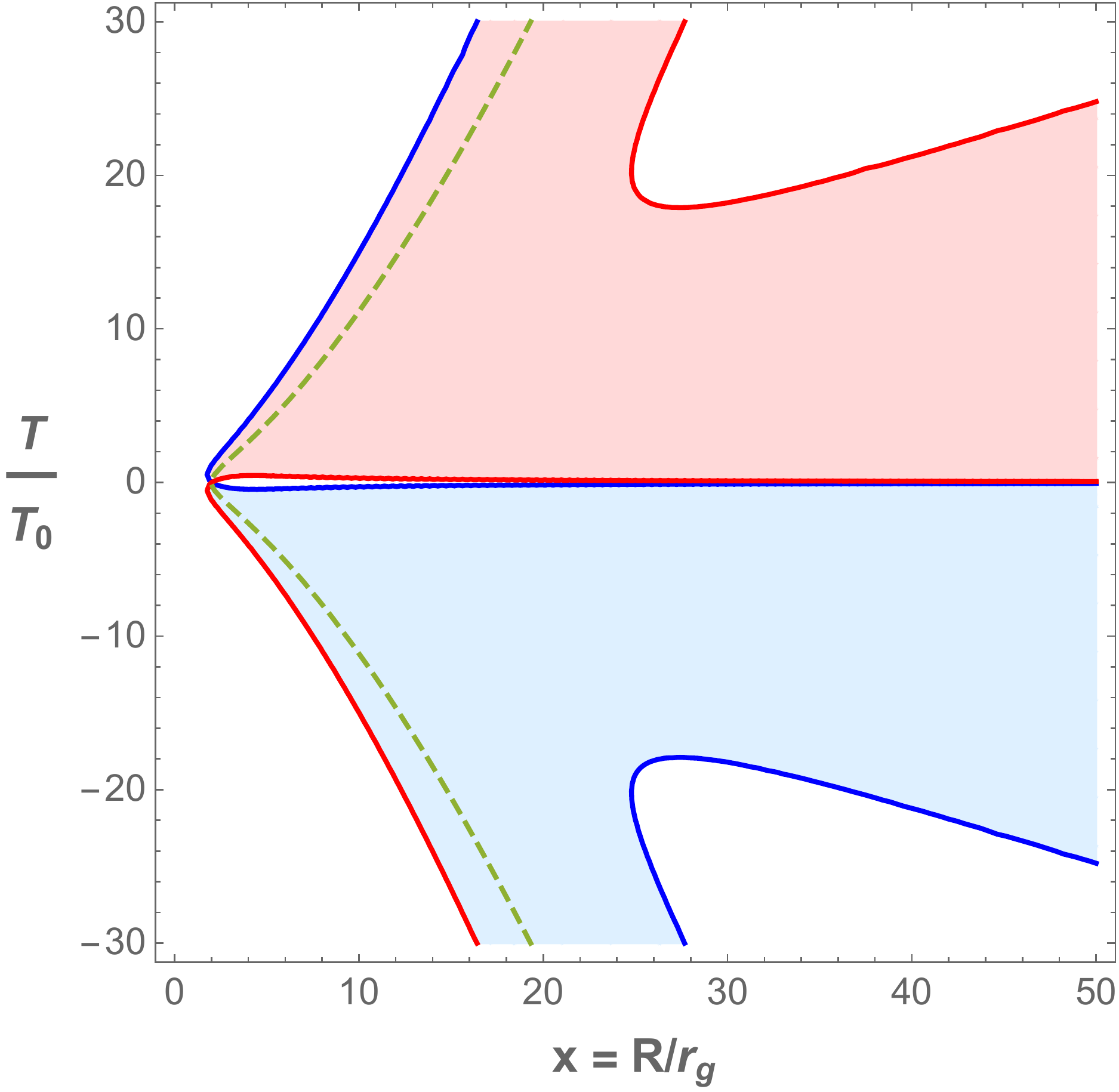}
\caption{\label{fig:1} The blue and red lines indicates the conditions $ \theta_{\mathrm{out}} = 0$ and  $ \theta_{\mathrm{in}}= 0$, respectively. The white zones point out the regular regions of the spacetime ($\theta_{\mathrm{in}} \theta_{\mathrm{out}} < 0$), the light pink zone the anti-trapped region ($\theta_{\mathrm{in}} \theta_{\mathrm{out}} > 0$, with $\theta_{\mathrm{out}} > 0$ and $\theta_{\mathrm{in}}  > 0$), and the light blue zone marks the trapped region ($\theta_{\mathrm{in}} \theta_{\mathrm{out}} > 0 $, with $\theta_{\mathrm{out}} < 0$ and $\theta_{\mathrm{in}}  < 0$). The dashed green line denotes the surface $R = 2 r_{g} a(T)$. Here, $T_{b} = 10^{-4}$ s  and $m_{0} = 10 \; M_{\odot}$.}
\end{figure}

We show in Figure \ref{fig:1} a plot of the trapping horizons and the different spacetime regions for the scale factor given by Eq. \eqref{scale-factor}. The blue line corresponds to $\theta_{\mathrm{out}} = 0$ whereas the red curve represents $\theta_{\mathrm{in}} = 0$. The dashed green line draws $R = 2 r_{g} a(T)$, where $r_{g} \equiv G \; m_0/c^{2}$ is the gravitational radius. The anti-trapped region is painted in light pink, the trapped region is in light blue, and the regular zones of the spacetime are in white.

As expected, in the limit $a(T) \rightarrow 1$ ($H(T) \rightarrow 0$), the line element \eqref{ds-areal-radius} reduces to the Schwarzschild line element
\begin{eqnarray}
 ds^{2} & = &  -c^2 \left(1 - \frac{2 G m_0}{c^2 R}\right) dT^2 + \frac{dR^2}{\left(1 - \frac{2 G m_0}{c^2 R}\right)} \nonumber \\
 & + &  R^2   \left(d\theta^{2} + \sin^{2}{\theta} d\phi^{2}\right).
 \end{eqnarray}
In Schwarzschild spacetime the coordinate $T$ is the proper time measured by an observer at rest in infinity, and $R$ is still the areal radius coordinate. The expressions for $\theta_{\mathrm{out}}$ and $\theta_{\mathrm{in}}$ become
\begin{eqnarray}
 \theta_{\mathrm{out}} & = &  \frac{2 \ c}{R}\left[ 1 - \frac{2 G m_0}{c^2 R} \right], \label{gout-sch}\\
 \theta_{\mathrm{in}} & = & \frac{- 2 \ c}{R}\left[ 1 - \frac{2 G m_0}{c^2 R} \right].\label{gin-sch}
 \end{eqnarray}
 
 We see that both $\theta_{\mathrm{out}} = \theta_{\mathrm{in}} = 0$ at $R = 2 r_{g}$, the event horizon of the black hole. The regions interior and exterior to the horizon are both regular:
 \begin{itemize}
 \item If $R > 2 \ r_{g}$, $\theta_{\mathrm{out}} > 0$ and $\theta_{\mathrm{in}} < 0 \;  \Rightarrow \; \theta_{\mathrm{out}}\theta_{\mathrm{in}} <0 $.
  \item If $R < 2 \ r_{g}$, $\theta_{\mathrm{out}} < 0$ and $\theta_{\mathrm{in}} > 0 \; \Rightarrow \; \theta_{\mathrm{out}}\theta_{\mathrm{in}} <0 $.
 \end{itemize}
 This result seems contrary to what we would expect for a Schwarzschild black hole; the region $R < 2 r_{g}$ should be trapped and not regular as we have just found\footnote{In Schwarzschild spacetime the line element in Painlev\'e-Gullstrand coordinates $(\bar{t},r,\theta,\phi)$ takes the form
 \begin{equation*}
 ds^{2}  =  - c^{2} \left(1 - \frac{2GM}{c^{2}r}\right) d\bar{t}^{2} + 2 \; c \sqrt{\frac{2GM}{c^{2}r}} d\bar{t} dr + dr^{2}+ r^{2} d\Omega^{2}. 
 \end{equation*}
 In these coordinates $\theta_{\mathrm{out}}$ and $\theta_{\mathrm{in}}$ read
 \begin{equation*}
  \theta_{\mathrm{out}}  =  \frac{2c}{r}\left(1 - \sqrt{\frac{2GM}{c^{2}r}}\right), 
 \end{equation*}
 \begin{equation*}
\theta_{\mathrm{in}}  = - \frac{2c}{r}\left(1 + \sqrt{\frac{2GM}{c^{2}r}}\right).   
 \end{equation*}
 We see that $\theta_{\mathrm{out}} \theta_{\mathrm{in}} < 0$ for $r> 2GM/c^{2}$ (regular spacetime region), and $\theta_{\mathrm{out}} \theta_{\mathrm{in}} > 0$, where $\theta_{\mathrm{in}}< 0$ and $\theta_{\mathrm{out}} < 0$ for $r< 2GM/c^{2}$ (trapped spacetime region)}.

 Since in the Schwarzschild limit ($a(T) \rightarrow 1$), the coordinates $\left(T,R,\theta,\phi\right)$, and thus the corresponding spacetime foliation, are inappropriate  to reveal the existence of a trapped region, we conclude that these same coordinates might not be adequate to analyze a much more complex spacetime such as the comoving generalized McVittie spacetime. Instead, Painlev\'e-Gullstrand coordinates are much better suited to our task, as we show next.

 
 \subsection{Comoving  generalized McVittie in Painlev\'e-Gullstrand coordinates}\label{CGMcVPG}
 
 The CGMcV line element in coordinates $(T, \tilde{r},\theta,\phi)$ can be rewritten as
  \begin{equation}\label{comoving}
 ds^{2}  =   a^2(T)\widetilde{ds}^2,
 \end{equation}
where
\begin{eqnarray}\label{conf-schw}
\widetilde{ds}^2 & = &  - \frac{c^{2}}{a^2(T)}  \left(1 -  \frac{2 G m_{0}}{ c^{2} \tilde{r}}\right) dT^{2}\\
& + & \left(1 -  \frac{2 G m_{0}}{ c^{2} \tilde{r}}\right)^{-1} {d\tilde{r}}^{2} + {\tilde{r}}^{2}\left(d\theta^{2} + \sin^{2}{\theta} d\phi^{2}\right).\nonumber
\end{eqnarray}

 The line element $\widetilde{ds}^2$ can be cast in a ``Schwarzschild-like form'' by defining a new coordinate $\tau$
 \begin{equation}
d\tau = \frac{1}{a(T)} dT.
\end{equation}
Then,  Eq. \eqref{conf-schw} takes the form
 \begin{eqnarray}\label{conf-schw-1}
\widetilde{ds}^2 & = &  - c^{2}  \left(1 -  \frac{2 G m_{0}}{ c^{2} \tilde{r}}\right) d\tau^{2}\\
& + & \left(1 -  \frac{2 G m_{0}}{ c^{2} \tilde{r}}\right)^{-1} {d\tilde{r}}^{2} + {\tilde{r}}^{2}\left(d\theta^{2} + \sin^{2}{\theta} d\phi^{2}\right).\nonumber
\end{eqnarray}

Though Schwarzschild coordinates are useful to describe certain properties of static black holes, they are not convenient to analyse the trapping horizons of a dynamical black hole. Instead, Painlev\'e-Gullstrand coordinates are well behaved at the horizons and as we will later show, essential to understand the casual structure of this spacetime.


 So, we perform a new coordinate transformation from  $(\tau, \tilde{r}, \theta,\phi)$ to  $(\tilde{t} ,\tilde{r}, \theta,\phi)$ where the coordinate $\tilde{t}$ is the Painlev\'e-Gullstrand (PG) time \citep{nil+06} \footnote{The relation between $\tau$ and $T$ is
 \begin{equation}
 \int \frac{1}{a(T)} dT = T   \;  _{2}F_{1}\left(\frac{1}{3},\frac{1}{2},\frac{3}{2};\frac{-T^{2}}{T_b}\right).  
 \end{equation}
Here, $_{2}F_{1}$ is a special function represented by the hypergeometric series. Notice that all our final results and plots are given in terms of the cosmic time $T$, so we do not need to compute  $a(\tilde{t},\tilde{r})$ explicitly.}:
\begin{eqnarray}
d\tilde{t} & = &  \frac{\partial \tilde{t}}{\partial{\tau}} d\tau + \frac{\partial \tilde{t}}{\partial \tilde{r}}   d\tilde{r} =   d\tau + \frac{1}{c \;f(\tilde{r})} d\tilde{r}, \label{t-pg}\\
\tilde{t} & = & \tau + \int \frac{1}{c \;f(\tilde{r})} d\tilde{r},\label{t-pg1}
\end{eqnarray}
 where
 \begin{equation}
 f(\tilde{r}) = \frac{ \left(1 -  \frac{2 G m_{0}}{ c^{2} \tilde{r}}\right)}{\sqrt{\frac{2 G m_{0}}{ c^{2} \tilde{r}}}}.  
 \end{equation}
 Substitution of Eq. \eqref{t-pg} into Eqs. \eqref{conf-schw-1} yields
 \begin{eqnarray}\label{schw-PG}
\widetilde{ds}^2 & = & - c^2    \left(1 -  \frac{2 G m_{0}}{ c^{2} \tilde{r}}\right) d\tilde{t}^2 + 2 \; c \sqrt{\frac{2Gm_0}{c^2 \tilde{r}}} d\tilde{t} d\tilde{r} + d\tilde{r}^2 \nonumber \\
& + &  {\tilde{r}}^{2}\left(d\theta^{2} + \sin^{2}{\theta} d\phi^{2}\right).
 \end{eqnarray}
 
 Notice that under the change of variables $T\rightarrow  \tau \rightarrow \tilde{t}$, the scale factor in the line element \eqref{comoving} is now a function of both $\tilde{t}$ and $\tilde{r}$, that is
 \begin{equation}
 a(T) \rightarrow a(T(\tau(\tilde{t}, \tilde{r}))).  
 \end{equation}
 
 In what follows, we denote the CGMcV metric $\tilde{g}_{ab}$ and the metric given by  \eqref{schw-PG} $g_{ab}$. Both metrics are expressed in terms of the PG coordinates $(\tilde{t}, \tilde{r},\theta,\phi)$
 \begin{equation}\label{rel-conf}
 \tilde{g}_{ab} = a^2(\tilde{t},\tilde{r})  g_{ab}.
 \end{equation}
 
Our next step is to compute the trapping horizons and determine the trajectories of the radial null geodesics of this spacetime. This is done in the following lines.

 \subsubsection{Trapping horizons}
 
 The outgoing (ingoing) radial null geodesic congruence of the metric  $g_{ab}$ have tangent fields \citep{nil+06}:
 \begin{eqnarray}
 l^{\mu} & = &  \left(\frac{1}{c},1- \sqrt{\frac{2Gm_0}{c^2 \tilde{r}}},0,0 \right), \label{out-radial}\\
  n^{\mu} & = & \left(\frac{1}{c},-1- \sqrt{\frac{2Gm_0}{c^2 \tilde{r}}},0,0 \right).\label{in-radial}
  \end{eqnarray}
We use these two vectors to calculate $\theta_{\mathrm{out}}$ and $\theta_{\mathrm{in}}$ following the definitions given by Eqs. \eqref{theta-out} and \eqref{theta-in}. We obtain
\begin{eqnarray}
 \theta_{l} & = &   \frac{2 }{r_g \; x}\left[f_1(x) + \frac{r_{g} x}{c} \frac{\dot{a}(\tilde{t},x)}{a(\tilde{t},x)} + f_{1}(x)r_{g} x \frac{a'(\tilde{t},x)}{a(\tilde{t},x)}\right], \nonumber\\
  \theta_{n} & = &   \frac{-2 }{r_g \; x}\left[f_2(x) - \frac{r_{g} x}{c} \frac{\dot{a}(\tilde{t},x)}{a(\tilde{t},x)} + f_{2}(x)r_{g} x \frac{a'(\tilde{t},x)}{a(\tilde{t},x)}\right] \nonumber,
\end{eqnarray}
 \begin{eqnarray}
   f_1(x) & = & 1- \sqrt{\frac{2}{x}}, \nonumber \\
   f_2(x) & = & 1+ \sqrt{\frac{2}{x}}.\nonumber
 \end{eqnarray}
The overdot indicates the derivative with respect to $\tilde{t}$ and the prime represents the derivative with respect to $x \equiv \tilde{r}/r_g$. In the limit $a \rightarrow 1$, we recover the expressions for the expansions in Schwarzschild spacetime. In terms of the scale factor given by \eqref{scale-factor} and the dimensionless cosmic time $\bar{T} = T/T_b$, $\theta_{l}$ and $\theta_{n}$ take the form
\begin{align}
  \theta_{l} & =  \frac{2 }{r_g \; x}\left[f_1(x) +   g(\bar{T},x) \left(1 + \frac{\sqrt{\frac{2}{x}}}{f_2(x)}\right)\right], \\
  \theta_{n} & =  \frac{- 2 }{r_g \; x}\left[f_2(x) +  g(\bar{T},x) \left(- 1 + \frac{\sqrt{\frac{2}{x}}}{f_1(x)}\right)\right], \\
  g(\bar{T},x) & = \frac{2 \ r_g \; x}{3 \ c} \frac{\bar{T}}{T_b} \frac{1}{\left(1 + \bar{T}^2\right)^{4/3}}.
  \end{align}

\begin{figure}[t]
\includegraphics[width=8cm]{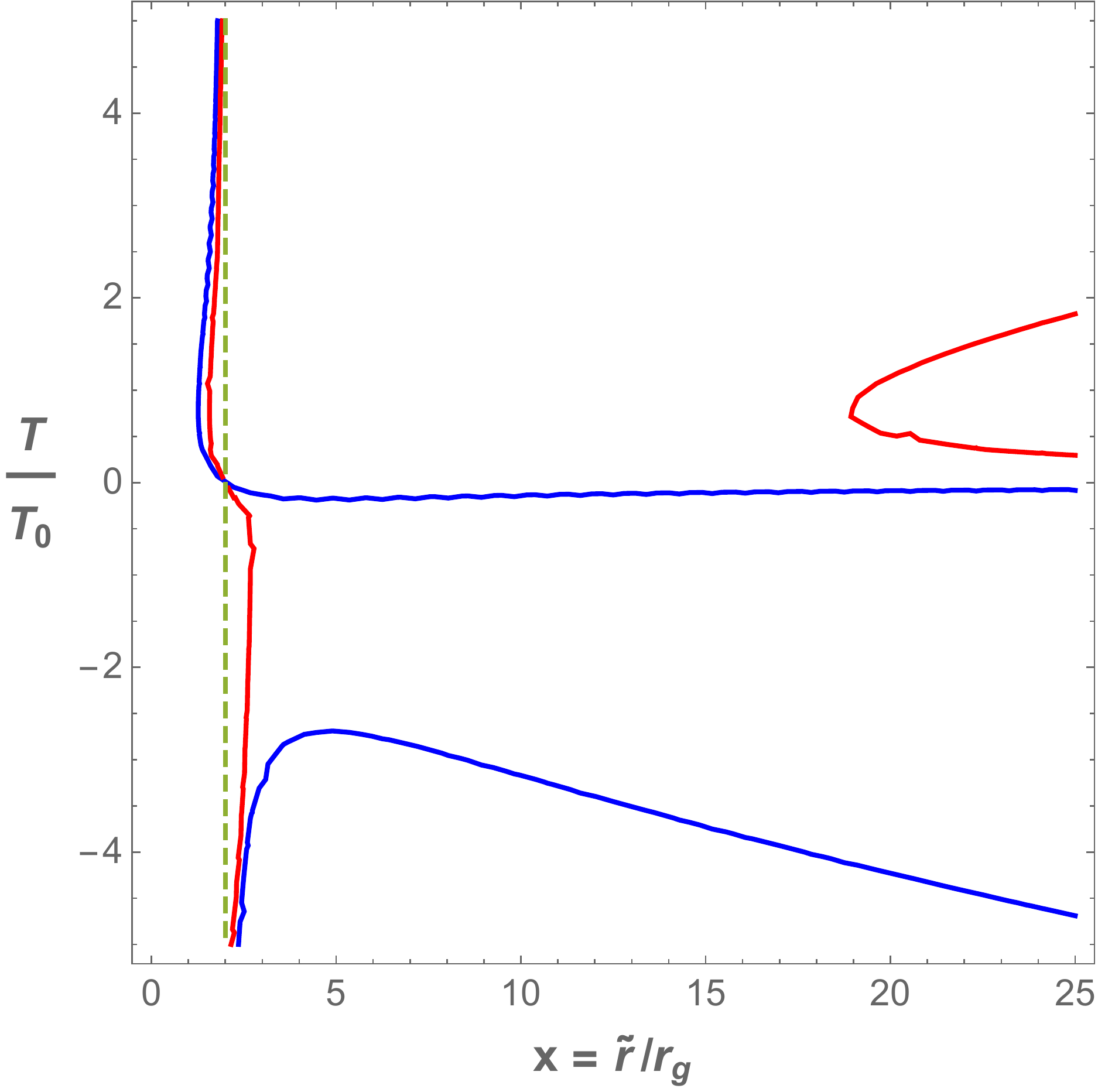}
\caption{\label{fig:2} The blue and red lines indicates the conditions $\theta_{\mathrm{out}} = 0$ and  $ \theta_{\mathrm{in}}= 0$, respectively. The dashed green line denotes the surface $\tilde{r} = 2 r_{g} $. Here, $T_{b} = 10^{-4}$ s  and $m_{0} = 10 \; M_{\odot}$.}
\end{figure}

\begin{figure*}[t]
\includegraphics[width=15cm]{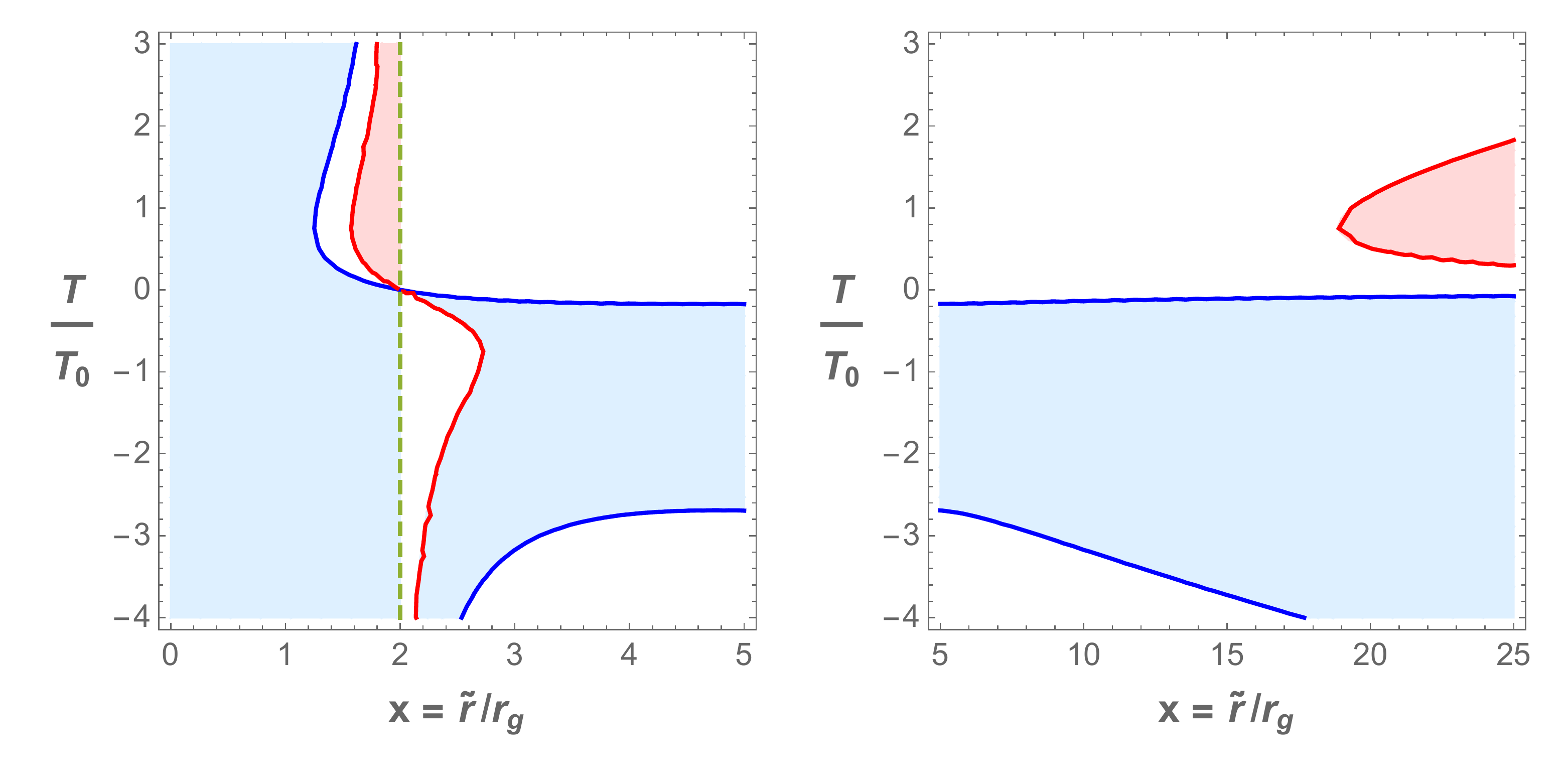}
\caption{\label{fig:3} The blue and red lines indicates the conditions $ \theta_{\mathrm{out}} = 0$ and  $ \theta_{\mathrm{in}}= 0$, respectively. The dashed green line denotes the surface $\tilde{r} = 2 r_{g} $. The trapped regions are painted in light blue, the anti-trapped are coloured in light pink and the regular zones are in white. Here, $T_{0} = 10^{-4}$ s  and $m_{0} = 10 \; M_{\odot}$.}
\end{figure*}

The equations $\theta_{\mathrm{out}} = 0$  (blue line) and  $ \theta_{\mathrm{in}}= 0$ (red line) in terms of the coordinates  $x$ and $T/T_b$ are plotted in Figure \ref{fig:2}.  The dashed green line denotes the surface $\tilde{r} = 2 r_{g} $. In Fig. \ref{fig:3} we show the trapped, anti-trapped and regular regions of the spacetime.

 

 \subsubsection{Radial null geodesics, light cones and Penrose diagrams}
 
\begin{figure}[t]
\includegraphics[width=8cm]{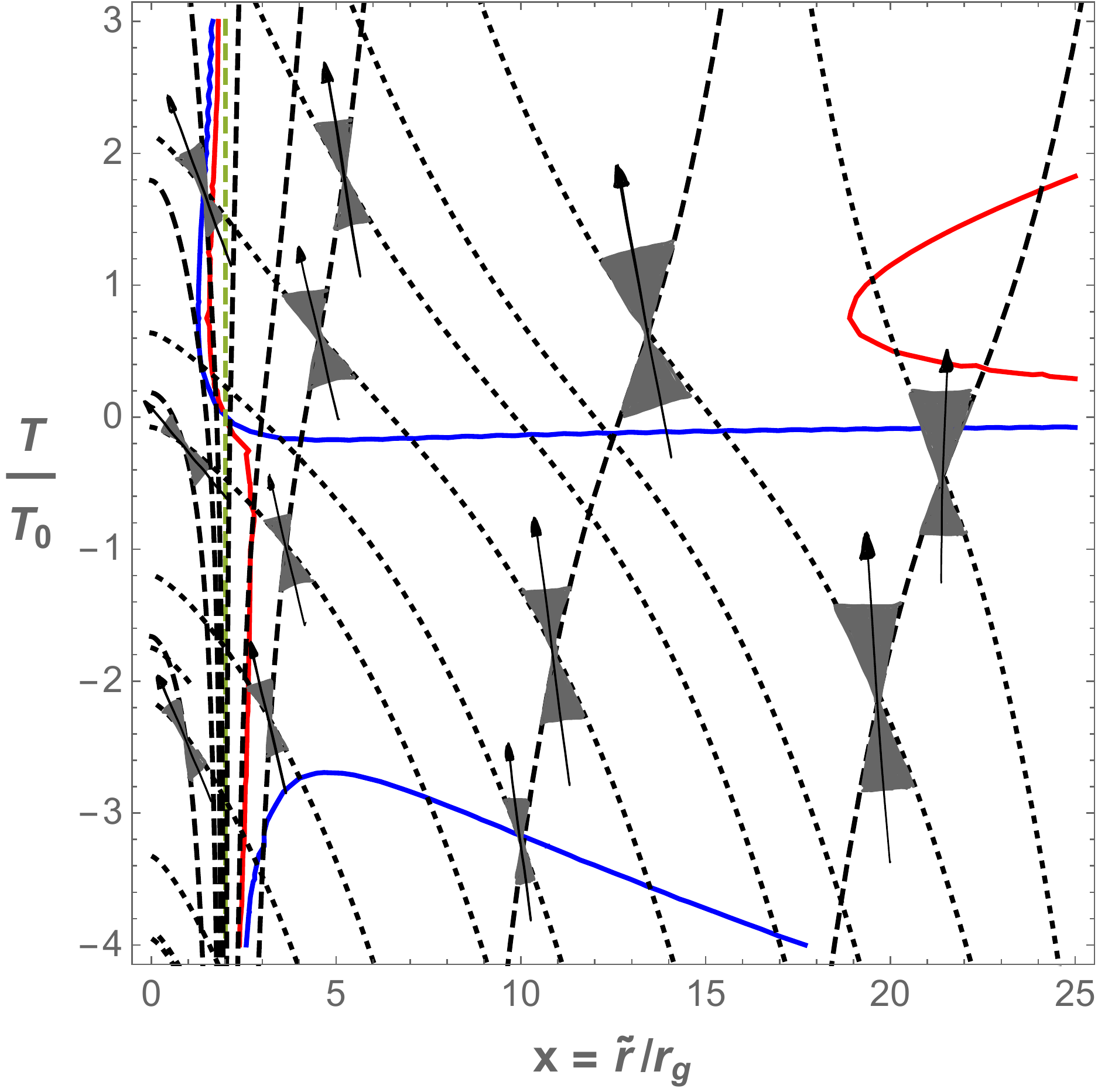}
\caption{\label{fig:4}The dotted (dashed) curves represent the null ingoing (outgoing) radial geodesics. The grey shadow regions show some light cones and the black arrow indicates the future direction. Here, $T_{b} = 10^{-4}$ s  and $m_{0} = 10 \; M_{\odot}$.}
\end{figure}

 We derive the equation for the outgoing and ingoing radial null geodesics by setting $\widetilde{ds}^2 =0$ and $d\theta = d\phi = 0$
 \begin{equation}
\left. \frac{dx}{d \bar{T}} \right \vert_{\pm}=  \frac{c \; T_b}{a(\bar{T}) r_{g}} \left( \pm 1 - \sqrt{\frac{2}{x}}\right),
 \end{equation}
where the ``$+$'' (``$-$'') corresponds to the outgoing (ingoing) case.  The trajectories of the radial null geodesics are determined by integrating the latter equation. The result is shown in Figure \ref{fig:4}. The dotted curves represent the null ingoing geodesics while the dashed curves the null outgoing ones. The grey shadow regions show some light cones and the black arrow indicates the local future direction.

The trajectories of ingoing radial null geodesics have a negative slope for all values of the cosmic time and for all values of the radial coordinate
\begin{equation}
\left. \frac{dx}{d \bar{T}} \right \vert_{-}=  - \frac{c \; T_b}{a(\bar{T}) r_{g}} \left(  1 + \sqrt{\frac{2}{x}}\right) < 0.
  \end{equation}
We can see that some of the ingoing null rays go through the surface $x = 2$ and end up at the singular surface $x = 0$. Eventually, all ingoing null rays terminate in the singularity. These geodesics can cross the surface $x = 2$ in only one way: from $x > 2$ to $x < 2$ since the region enclosed by $x = 2$ is trapped.

Outgoing null geodesics are expanding in the region $x >2$ for all values of the cosmic time. As they get closer to $x = 2$, the slope of the trajectories becomes smaller and in the limit $x \rightarrow 2$
\begin{equation}
\lim_{x \rightarrow 2}  \left. \frac{dx}{d \bar{T}} \right \vert_{+} = \lim_{x \rightarrow 2} \frac{c \; T_b}{a(\bar{T}) r_{g}} \left( 1 - \sqrt{\frac{2}{x}}\right) \rightarrow 0.
\end{equation}
In the trapped region ($x < 2$), the slope of outgoing null rays is negative and these geodesics are interrupted at the singularity.

\begin{figure}[t]
\includegraphics[width=8cm]{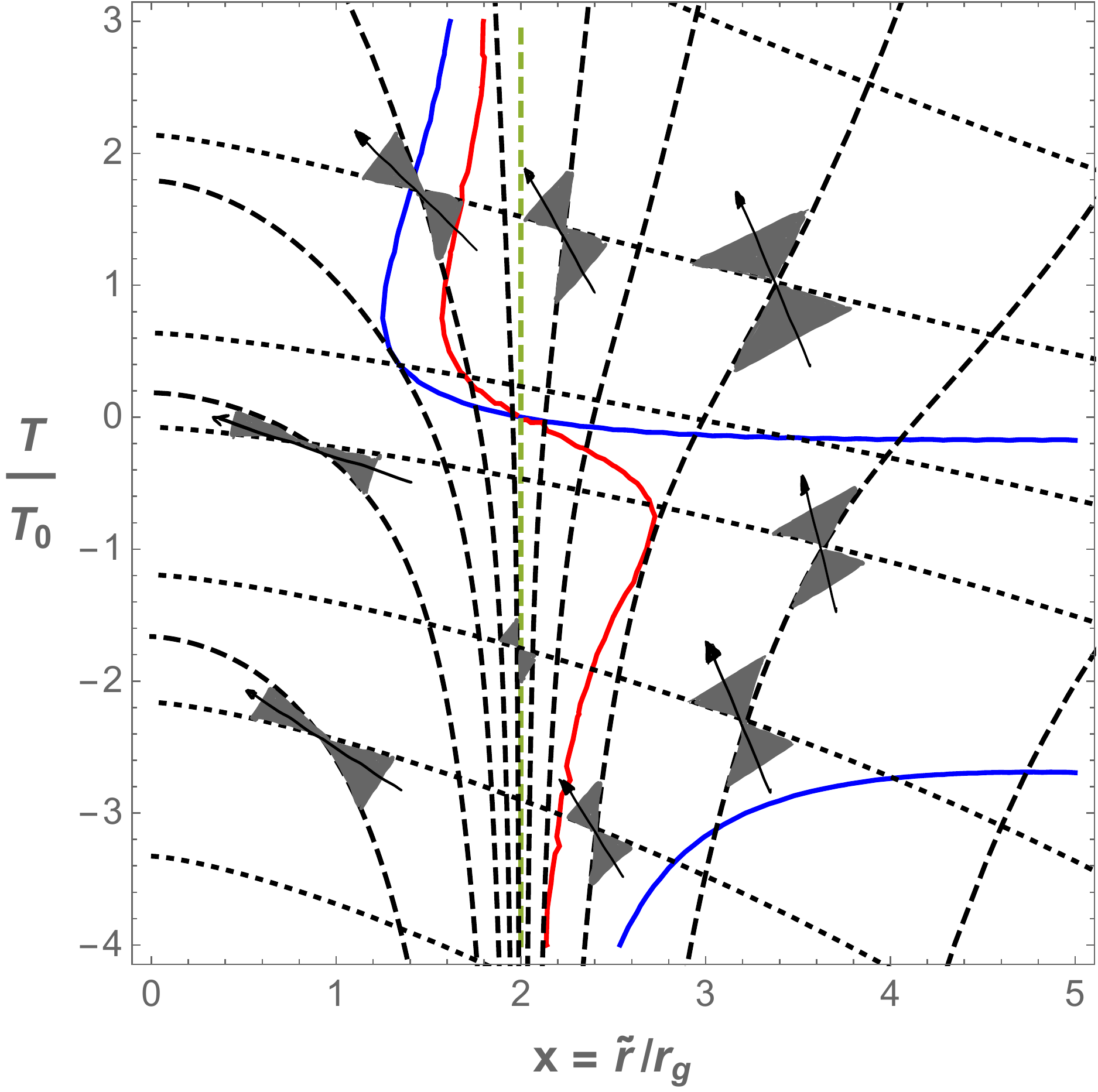}
\caption{\label{fig:5} Light cone structure close to the surface $x = 2$. The dotted (dashed) curves represent the null ingoing (outgoing) radial geodesics. The grey shadow regions show some light cones and the black arrow indicates the future direction. Here, $T_{b} = 10^{-4}$ s  and $m_{0} = 10 \; M_{\odot}$.}
\end{figure}

In Figure \ref{fig:5}, we offer a zoom into the region near the bounce close to $x = 2$. We see that all those particles that cross the surface $x = 2$ have in their local future the singularity at $x = 0$. The light cone structure makes evident that the surface $x =2$ acts as a one way membrane behaving like an event horizon that is present in all the cosmological epochs of the universe (contraction, bounce and expansion). Thus, we conclude that the comoving generalized McVittie spacetime in a bouncing cosmological model includes a dynamical black hole at all times

\begin{figure}[t]
\includegraphics[width=8cm]{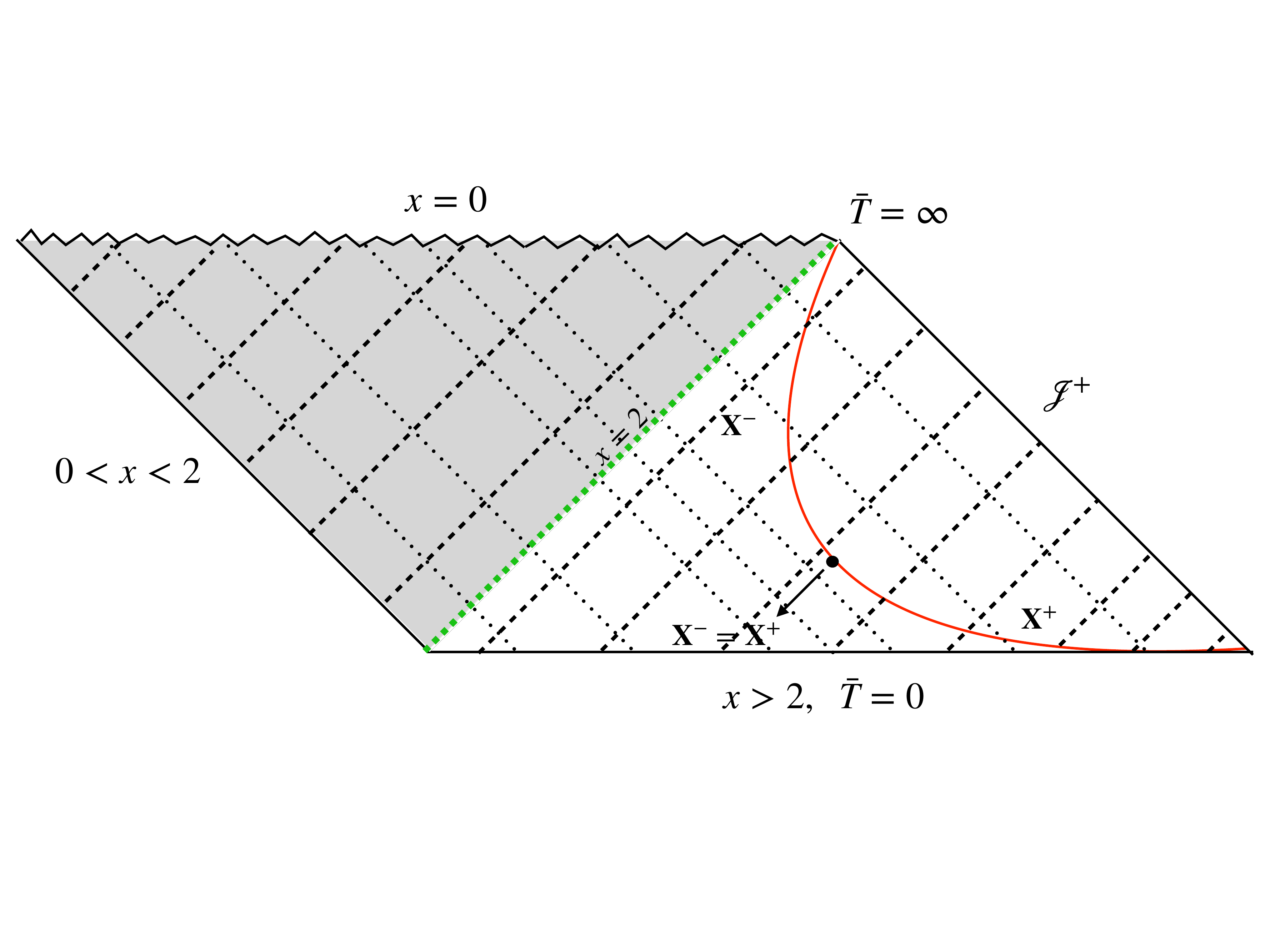}
\caption{\label{fig:6} Qualitative Penrose diagram of the CGMcV spacetime
for a bouncing cosmological model in the expanding
region ($t > 0$). The dotted lines represent null ingoing
geodesics while the dashed lines null outgoing geodesics. The grey shadow region corresponds to the black hole interior. Here, $\mathcal{J}^{+}$ represents the future null infinity.}
\end{figure}

\begin{figure}[t]
\includegraphics[width=8cm]{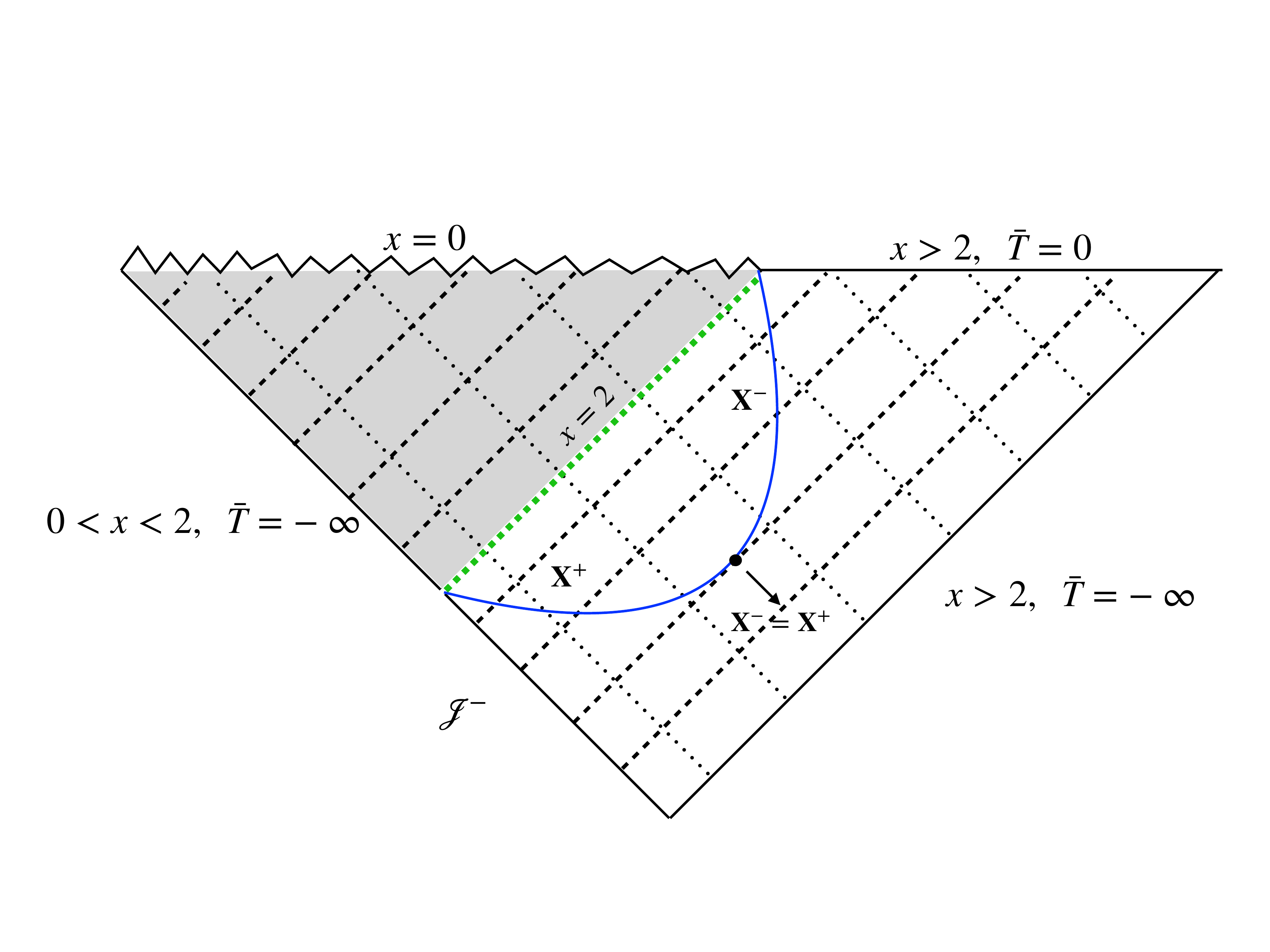}
\caption{\label{fig:7} Qualitative Penrose diagram of the CGMcV spacetime
for a bouncing cosmological model in the contracting
region ($t < 0$). The dotted lines represent null ingoing
geodesics while the dashed lines null outgoing geodesics. The grey shadow region corresponds to the black hole interior. Here, $\mathcal{J}^{-}$ represents the past null infinity.}
\end{figure}

To conclude our analysis, we show in Figures \ref{fig:6} and \ref{fig:7} qualitative Penrose diagrams of the  CGMcV spacetime in the expanding and contracting phase, respectively. The dotted lines represent the trajectories of ingoing null geodesics and the dashed lines the trajectories of outgoing null geodesics. In Fig. \ref{fig:6}, the red curve displays the inner ($X_{-}$) and outer ($X_{+}$) trapping horizons, while in Fig. \ref{fig:7} trapping horizons are shown by a blue curve. In both diagrams, the dashed green line corresponds to the surface $x = 2$, and the grey shadow region corresponds to the black hole interior. As usual, $\mathcal{J^{+}}$ ($\mathcal{J^{-}}$) denotes the future (past) null infinity.

In the both the expanding and contracting phase, all ingoing geodesics, regardless of the initial conditions, reach the surface $x =2$ and end up in the singularity. Outgoing null geodesics in the contracting phase and outside the trapped region ($0 < x < 2$, $t$ finite) cross the bounce and propagate reaching $\mathcal{J^{+}}$, the future null infinity. All the outgoing null geodesics in the trapped region finish in the singular surface $x =0$. The Penrose diagrams make clear that no null geodesic in the region $0 < x < 2$, $t$ finite can cross the surface $x = 2$ and escape to the future null infinity. The surface $x =2$ is the boundary between two causally disconnected spacetime regions during the phases of contraction, bounce and expansion of the universe. 

In terms of the areal radius coordinate $R = a(T) \tilde{r}$, the radius of the surface that encloses the black hole is
\begin{equation}
R(T) =  2 \frac{G m_0}{c^2} a(T),  
\end{equation}
and the corresponding Weyl part of the MSH energy is
\begin{equation}\label{mass}
M(T) =   M_0 \; a(T).
\end{equation}
We thus see that size of the black hole diminishes or grows according to the cosmic evolution of the universe. 

\section{Discussion and conclusions}

In this work we have analyzed the causal structure of the comoving  generalized McVittie spacetime in a bouncing cosmological model. We have computed the trapping horizons, the trajectories of radial  ingoing and outgoing null geodesics, and constructed the corresponding Penrose diagram. Our main result is that the solution represents a dynamical black hole that exists at all epochs of the bouncing cosmological model. 

In the contracting phase, the area of the black hole horizon decreases reaching its minimum value at the bounce. Once in the expanding phase, its size begins to increase proportionally to the square of the scale factor. This is not the case in the McVittie spacetime \cite{per+21a,per+21b}, where the black hole horizon merges with the cosmological horizon in the contracting phase before reaching the bounce. Such a solution, however, does not take into account the coupling of the black hole with the cosmic dynamics, as in the present work. Thus, we think that the results here presented correspond to a much more realistic scenario. 

If black holes survive  through a cosmological bounce, they might play an important role in the subsequent expanding phase of the universe. For instance, these surviving black holes might contribute to a fraction of the total dark matter component or provide the seeds for the formation of galaxies \cite{car+18,car+20}. 

On the other hand, supermassive black holes (SMBHs) are usually considered as the seeds for galaxy formation,  but their origin remains an unsolved problem. Recent observations have revealed the existence of a population of SMBHs with masses of $10^8-10^{10} \; M_{\odot}$ powering luminous quasars at $z >6$, when the universe was less than one billion years old \cite{yan+20,wan+21,vol+21}. There is no established mechanism that could explain how these black holes acquired such hulking masses at an early stage of the cosmic evolution.

In the context of the model presented in this work, if we assume that at the time of the bounce $T = 0$, the surviving black hole has a mass \footnote{Limits on the $µ$-distortion in the CMB due to
the dissipation of fluctuations before decoupling exclude primordial black holes larger than $10^{5} \; M_{\odot}$.} of $M_0 = 10^{4} \; M_{\odot}$ \cite{car+18,car+20}, then one billion years after the bounce only the the expansion of the universe would increase the gravitational mass of the black hole (see Eq. \eqref{mass}) up to $10^{9} \; M_{\odot}$, reaching the values observed in the most distant quasars\footnote{We suppose that $a_{b} = 2.16 \times 10^{-5}$ in Eq. \eqref{scale-factor}. This value is in accordance with the restriction $a_{b} >> 10^{-38}$  \cite{Frion2020}.}. 

The results here presented open the possibility for further investigations. These include the construction of a cosmological model where a black hole population survives the bounce and produce perturbations in the distribution of cosmological fluid in the expanding phase, the interaction of black holes during the bounce, and the gravitational waves produced by these black holes. We shall explore some of these issues in future works.

\begin{acknowledgments}
The authors are grateful to Santiago E. Perez Bergliaffa for many insightful comments on this article. This work was supported by grant PIP 0554 (CONICET) and PIP 2021-1639 (CONICET). GER acknowledges the support from the Spanish Ministerio de Ciencia e Innovaci\'on (MICINN) under grant PID2019-105510GBC31 and through the Center of Excellence Mara de Maeztu 2020-2023 award to the ICCUB (CEX2019-000918-M). We are very grateful to an anonymous referee for her/his careful analysis that help  to improve this article.
\end{acknowledgments}




 \bibliography{apssamp}

\end{document}